\documentclass[12pt]{iopart}

\usepackage{graphicx}

\usepackage{epstopdf}
\begin{document}

\title[Charged particle directed flow in Pb-Pb collisions measured with ALICE]
{Charged particle directed flow in Pb-Pb collisions at $\sqrt{s_{\rm NN}} = $2.76 TeV measured with ALICE at the LHC}

\author{}

\address{{\bf Ilya Selyuzhenkov}$^{*}$ for the ALICE Collaboration
\\$^{*}$Research Division and ExtreMe Matter Institute EMMI,\\
GSI Helmholtzzentrum f\"ur Schwerionenforschung, Darmstadt, Germany}
\ead{ilya.selyuzhenkov@gmail.com}
\begin{abstract}
Charged particle directed flow at midrapidity, $|\eta|<0.8$, and forward rapidity, $1.7 < |\eta|<5.1$, 
is measured in Pb-Pb collisions at $\sqrt{s_{\rm NN}} = $~2.76~TeV
with ALICE at the LHC.
Directed flow is reported as a function of collision centrality,
charged particle transverse momentum, and pseudo-rapidity.
Results are compared to measurements at RHIC
and recent model calculations for LHC energies.
\end{abstract}

\noindent{\bf Introduction.}~Azimuthal anisotropic flow is a key observable indicating collectivity
among particles produced in non-central heavy ion collisions.
Directed flow is characterized by the first harmonic coefficient, $v_1$,
in the Fourier decomposition of the particle azimuthal distribution
with respect to the collision reaction plane.
It develops at a very early stage of the collision and thus
is sensitive to the properties and the equation of state of
the hot and dense matter produced in nucleus-nucleus collisions.
Among features of charged particle $v_1$ measured at RHIC energies
\cite{Abelev:2008jga,Back:2005pc}
are negative slope at midrapidity and absence of the so-called
wiggle structure predicted by RQMD model calculations \cite{Snellings:1999bt}.
Recent calculations within quark-gluon string model with
parton rearrangement \cite{Bleibel:2007se},
and fluid dynamical prediction \cite{Csernai:2011gg}
suggest much stronger signal at LHC energies with
a positive slope of $v_1$ vs. rapidity.
We report on charged particle $v_1$ measured in a wide range of rapidity
in Pb-Pb collisions at $\sqrt{s_{\rm NN}} = $~2.76~TeV.
Results are compared to RHIC measurements
and recent model calculations for LHC energies.

\noindent{\bf Data analysis.}~Minimum bias trigger and standard ALICE event selection criteria
for Pb-Pb collisions \cite{Aamodt:2010pa} were applied in the analysis.
Only events within  \mbox{0-80\%} centrality range and
reconstructed collision vertex within 10 cm from the centre of
the Time Projection Chamber (TPC) \cite{2004:ALICE-PPR1}
were selected for the analysis (about 8 million events total).
A transverse momentum cut of $p_{\rm t}>0.15$ GeV/$c$ and pseudo-rapidity cut of $|\eta|<0.8$
are imposed on all charged particle tracks reconstructed with TPC.
Charged particle multiplicity at forward rapidity is measured by the pair
of forward scintillator arrays (VZERO) \cite{2004:ALICE-PPR1}, which
covers rapidity range of $-3.7<\eta<-1.7$ and $2.8<\eta<5.1$
and has the azimuthal segmentation required for anisotropic flow measurement.

\begin{figure}[ht]
\begin{center}
\includegraphics[width=.5\textwidth]{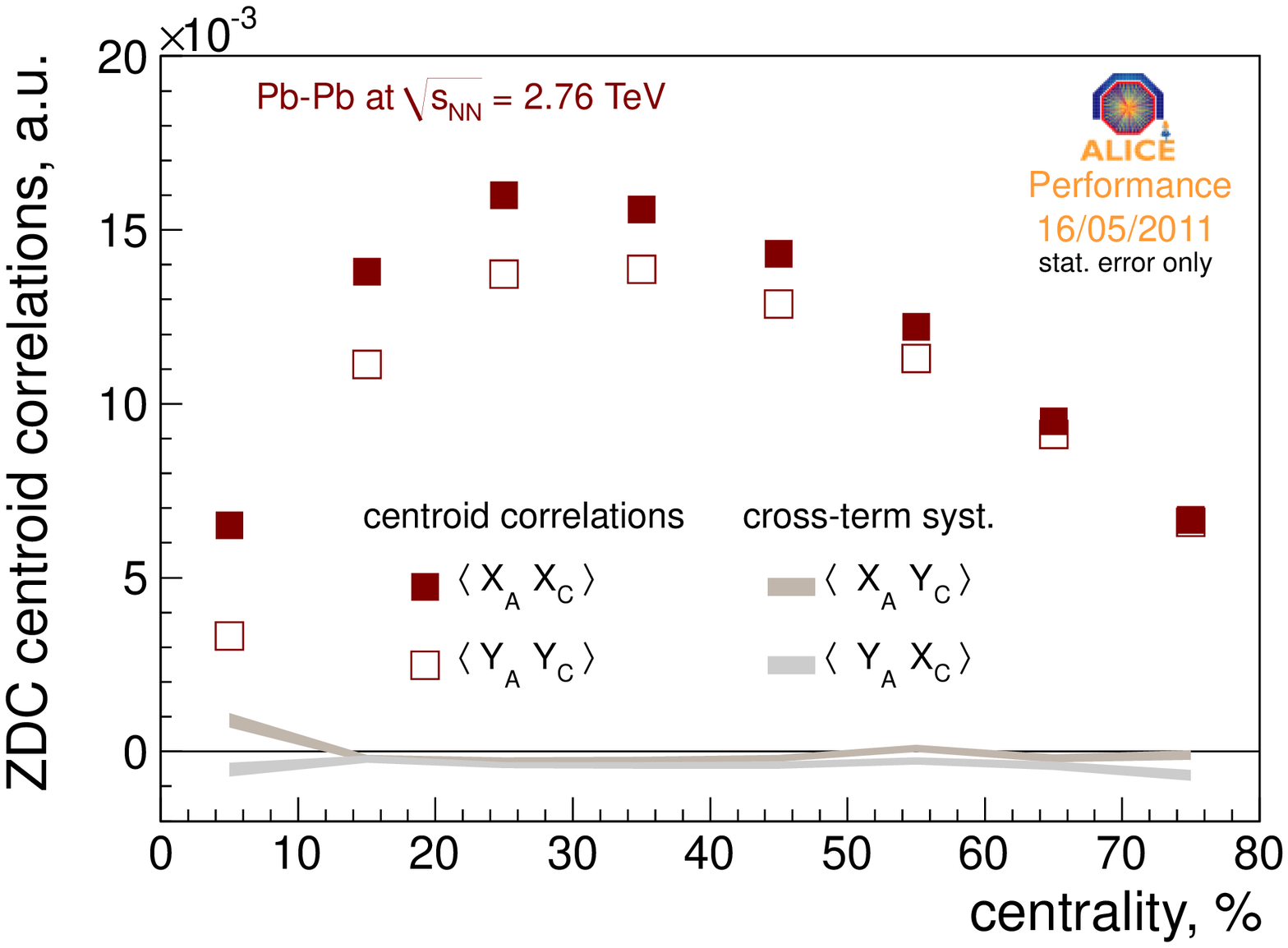}%
\includegraphics[width=.5\textwidth]{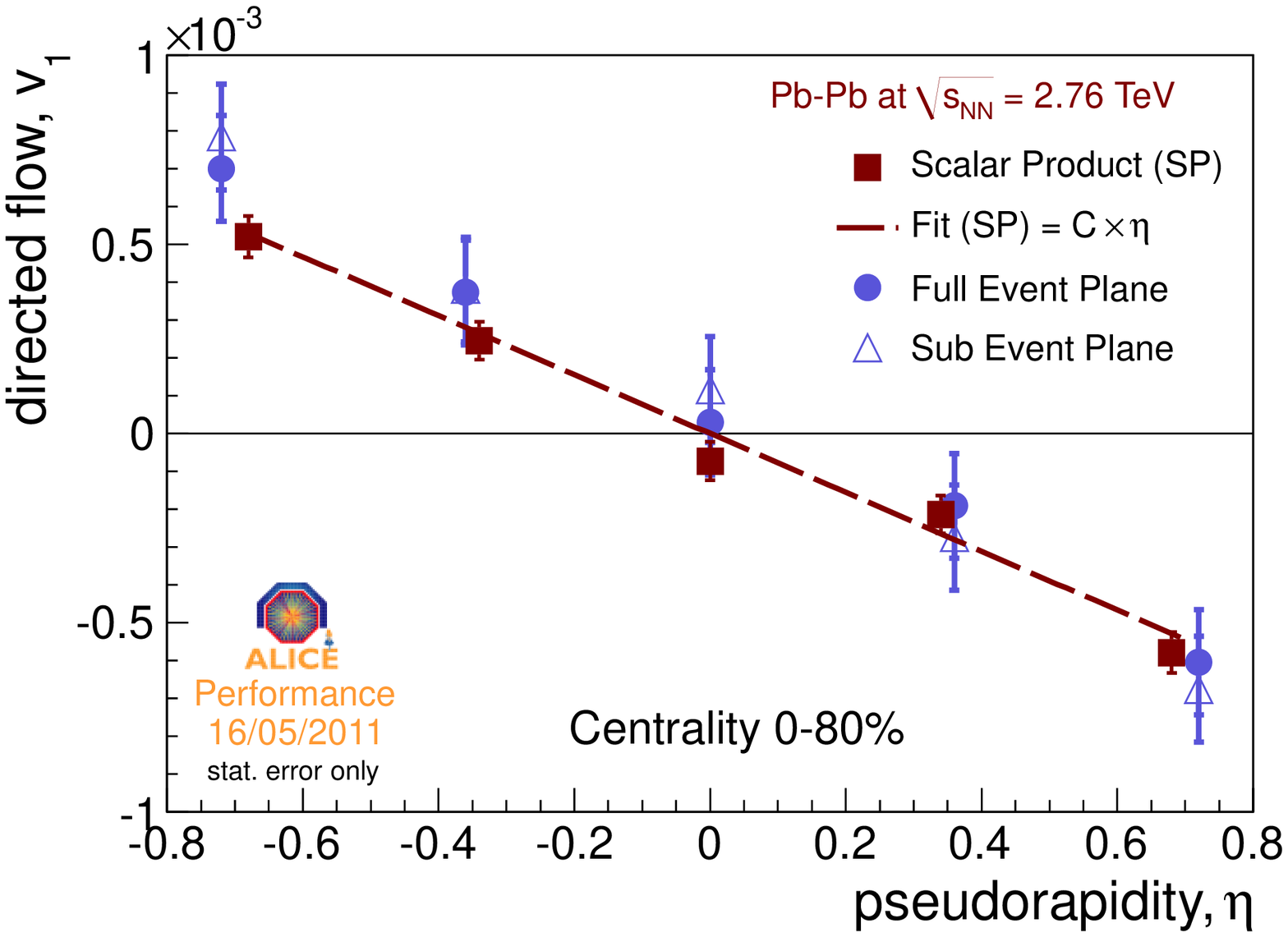}
{\mbox{}\\\vspace{-0.9cm}
\hspace{1cm}\mbox{~} \bf (a)
\hspace{+6.7cm}\mbox{~} \bf (b)}
{\mbox{}\\\vspace{-0.2cm}}
\caption{(color online) (a) ZDC centroid correlation after recentering;
(b) $v_1$ vs. pseudorapidity: comparison of the scalar product and event plane methods.}
\label{fig:1}
\end{center}
\end{figure}

The orientation of the collision reaction plane is reconstructed by the pair of
Zero Degree Calorimeters (ZDC) \cite{2004:ALICE-PPR1}.
Located 114 meters apart from the interaction point, ZDCs
are sensitive to neutron spectators at beam rapidity, $|\eta| \sim 8.8$.
Each ZDC, A-side ($\eta>0$) and C-side ($\eta<0$), has a $2\times 2$ tower geometry.
Event-by-event spectator deflection is estimated from
ZDC centroid shifts, ${\bf Q}$, defined by energy, $E_i$,
weighted mean of ZDC tower centers, ${\bf r}_i=(x_i,y_i)$
\begin{equation}
(a) ~~{\bf Q} = (X,~Y)  = {\sum\limits_{i=1}^{4}
{\bf r}_i E_i}/{\sum\limits_{i=1}^{4} E_i}
~~~~(b)~~{\bf Q}'= {\bf Q}-\langle {\bf Q}\rangle.
\label{equation:1}
\end{equation}
To correct for the time dependent variation of the beam crossing position
and event-by-event spread of the collision vertex with respect to the centre of TPC
we perform the recentering procedure defined by Eq.~\ref{equation:1}(b).
Recentering (subtracting the average centroid position
$\langle {\bf Q}\rangle$) is performed as a function of time, 
collision centrality, and transverse position of the collision vertex.
Only after recentering we observe consistent correlation between shifts
in the same direction for A and C sides ZDCs (Fig. \ref{fig:1}(a)).
Non-zero correlation for $\langle X_{\rm A} X_{\rm C}\rangle$ and $\langle Y_{\rm A} Y_{\rm C}\rangle$
reflects directed flow of spectators.
No correlation between centroid shifts in orthogonal directions,
$\langle X_{\rm A} Y_{\rm C}\rangle$ and $\langle Y_{\rm A} X_{\rm C}\rangle$,
indicates that recentering removes correlations
originating from the beam parameter variation.

The scalar product and the event plane methods are used for the $v_1$ measurement.
In the scalar product method, the directed flow is quantified by correlating ZDC centroid shifts
with the first harmonic azimuthal asymmetry of charged particles
\begin{equation}
v_1\{{\rm SP}\}= \sqrt{2}~{\langle \cos\phi ~X_{\rm C}\rangle}/{\sqrt{\langle X_{\rm A} ~X_{\rm C}\rangle}} 
~= ~\sqrt{2}~{\langle \sin\phi~ Y_{\rm C}\rangle}/{\sqrt{\langle Y_{\rm A} ~Y_{\rm C}\rangle}}.
\label{Equation:SP}
\end{equation}
Here $\phi$ is either charged particle azimuthal angle of TPC tracks
or multiplicity weighted azimuthal position of tiles in VZERO scintillator arrays.
No correction is applied to account for non-primary tracks pointing to VZERO.
In the event plane method centroid shifts define
the full, $\Psi_{\rm Full} = {\rm atan2}\{Y_{\rm A}+Y_{\rm C},X_{\rm A}+X_{\rm C}\}$,
and subevent, $\Psi_{\rm A,C} = {\rm atan2}\{Y_{\rm A,C},X_{\rm A,C}\}$,
event plane angles, which are later used to calculated $v_1$ according to equation
\begin{equation}
v_1\{\rm EP\} = {\langle \cos(\phi -\Psi_{\rm Full})\rangle}/
{\sqrt{2~\langle \cos(\Psi_{\rm A}-\Psi_{\rm C})\rangle}}.
\label{Equation:EP}
\end{equation}
The absolute sign of $v_1$ in Eqs.~(\ref{Equation:SP}, \ref{Equation:EP}) is fixed by the same
convention as used at RHIC, i.e.~spectators with $\eta>0$ possess a positive $v_1$.
Figure \ref{fig:1}(b) shows consistency between results from two different methods.
$v_1\{\rm SP\}$ is used to measure $v_1$ at midrapidity with TPC tracks,
while $v_1\{\rm EP\}$ is used to extract $v_1$ at forward rapidity with VZERO detectors.

\begin{figure}[ht]
\begin{center}
\includegraphics[width=.5\textwidth]{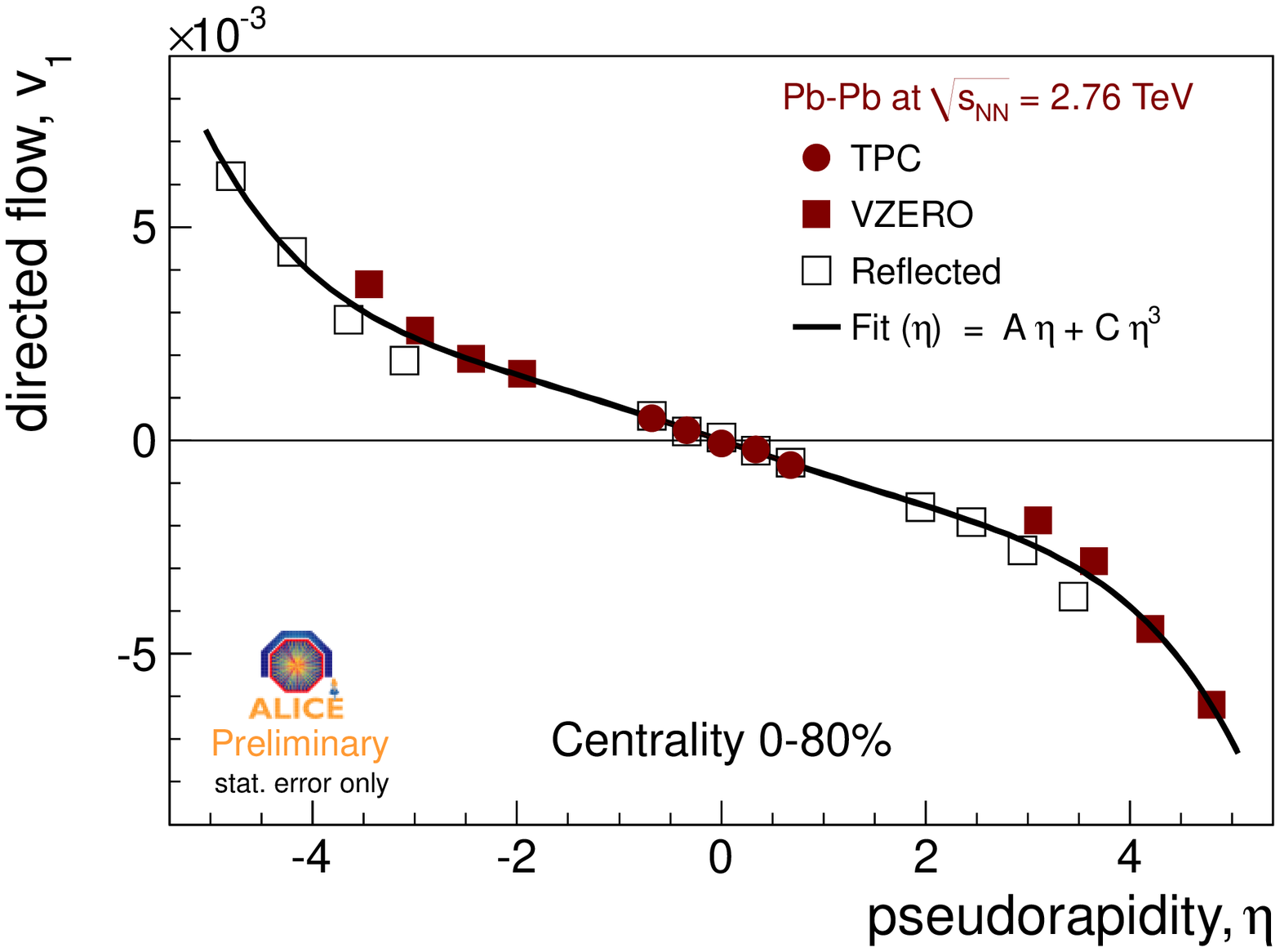}%
\includegraphics[width=.5\textwidth]{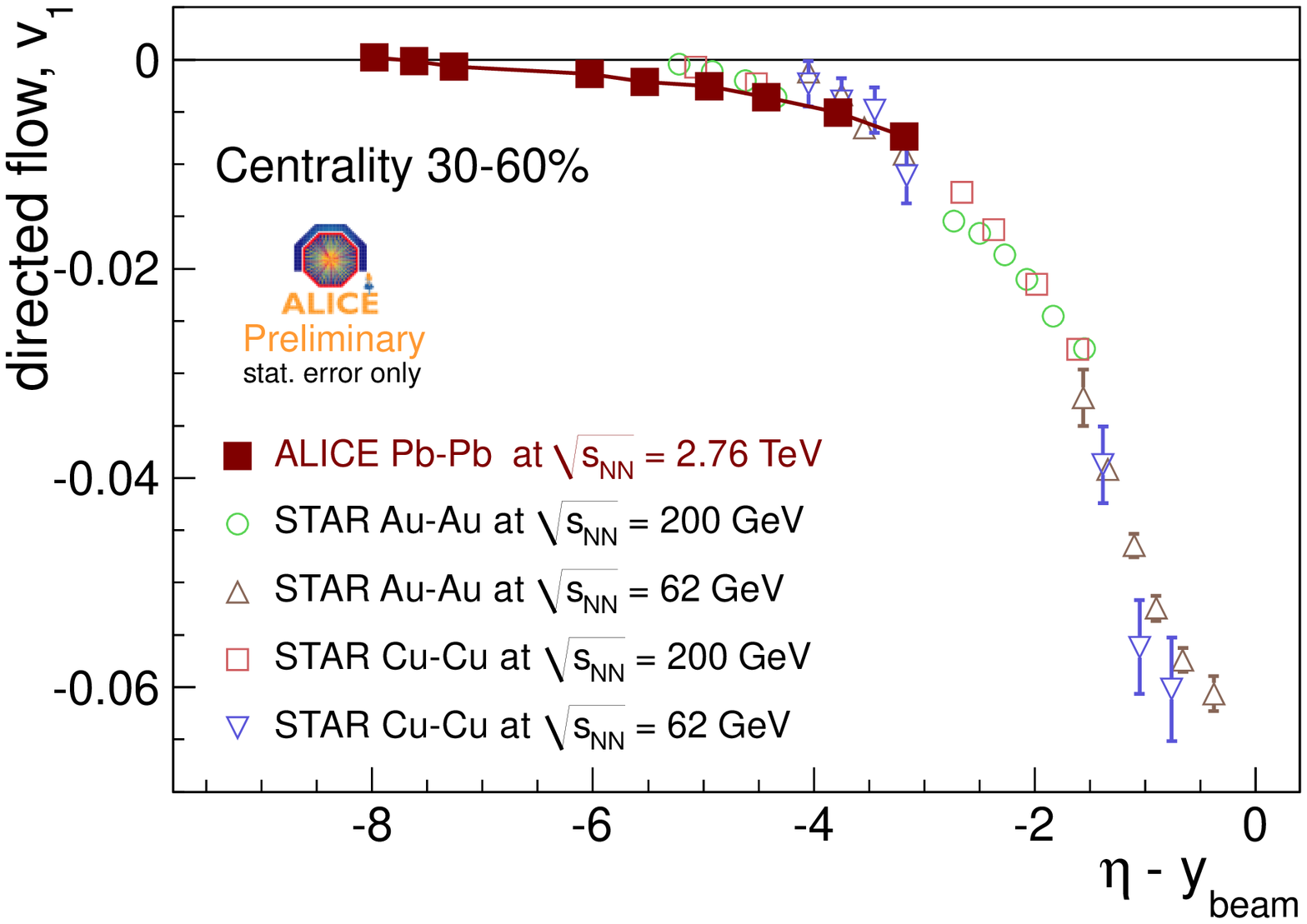}
{\mbox{}\\\vspace{-1.1cm}
\hspace{1cm}\mbox{~} \bf (a)
\hspace{+6.7cm}\mbox{~} \bf (b)}
{\mbox{}\\\vspace{-0.cm}}
\includegraphics[width=.5\textwidth]{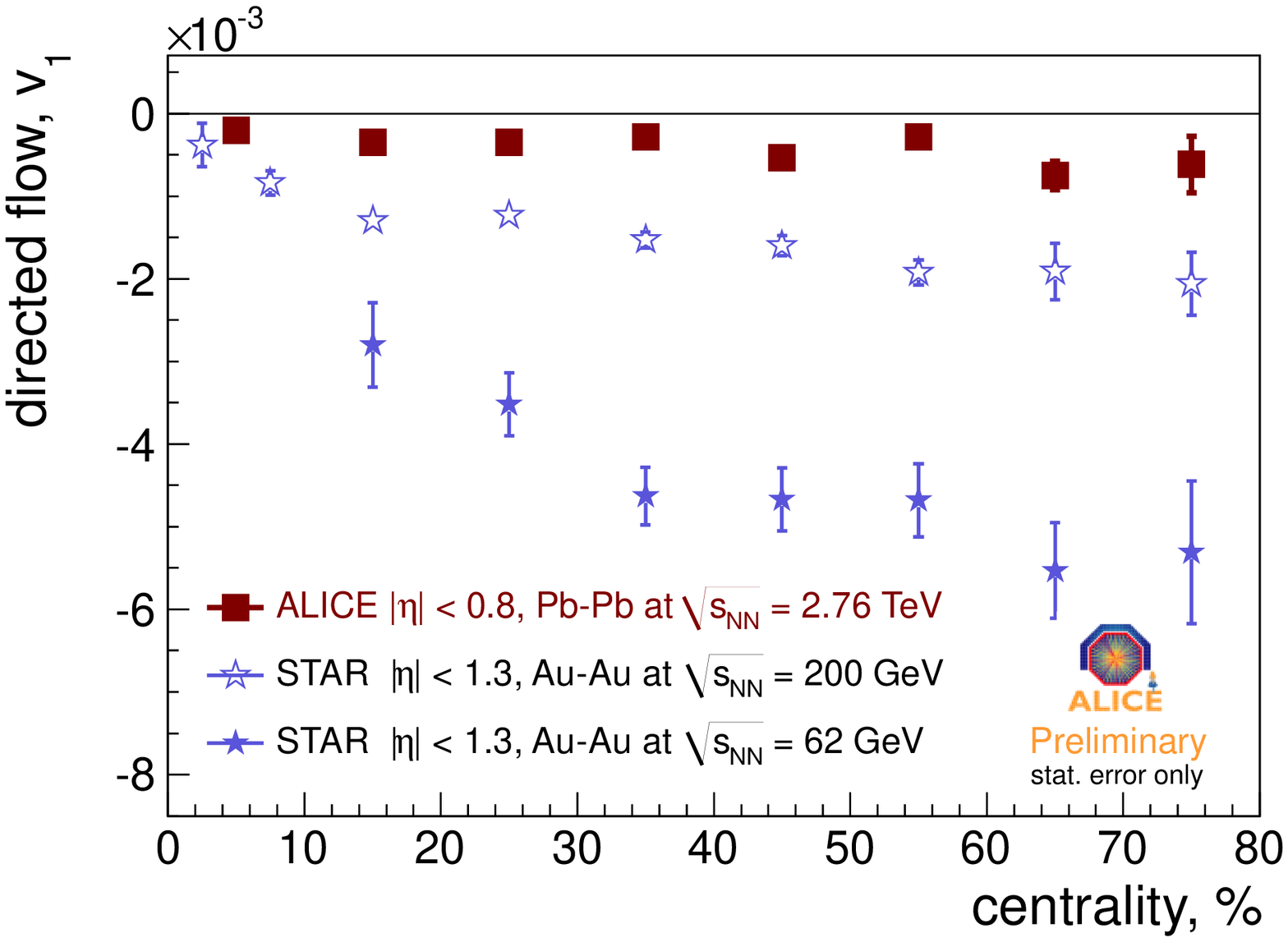}%
\includegraphics[width=.5\textwidth]{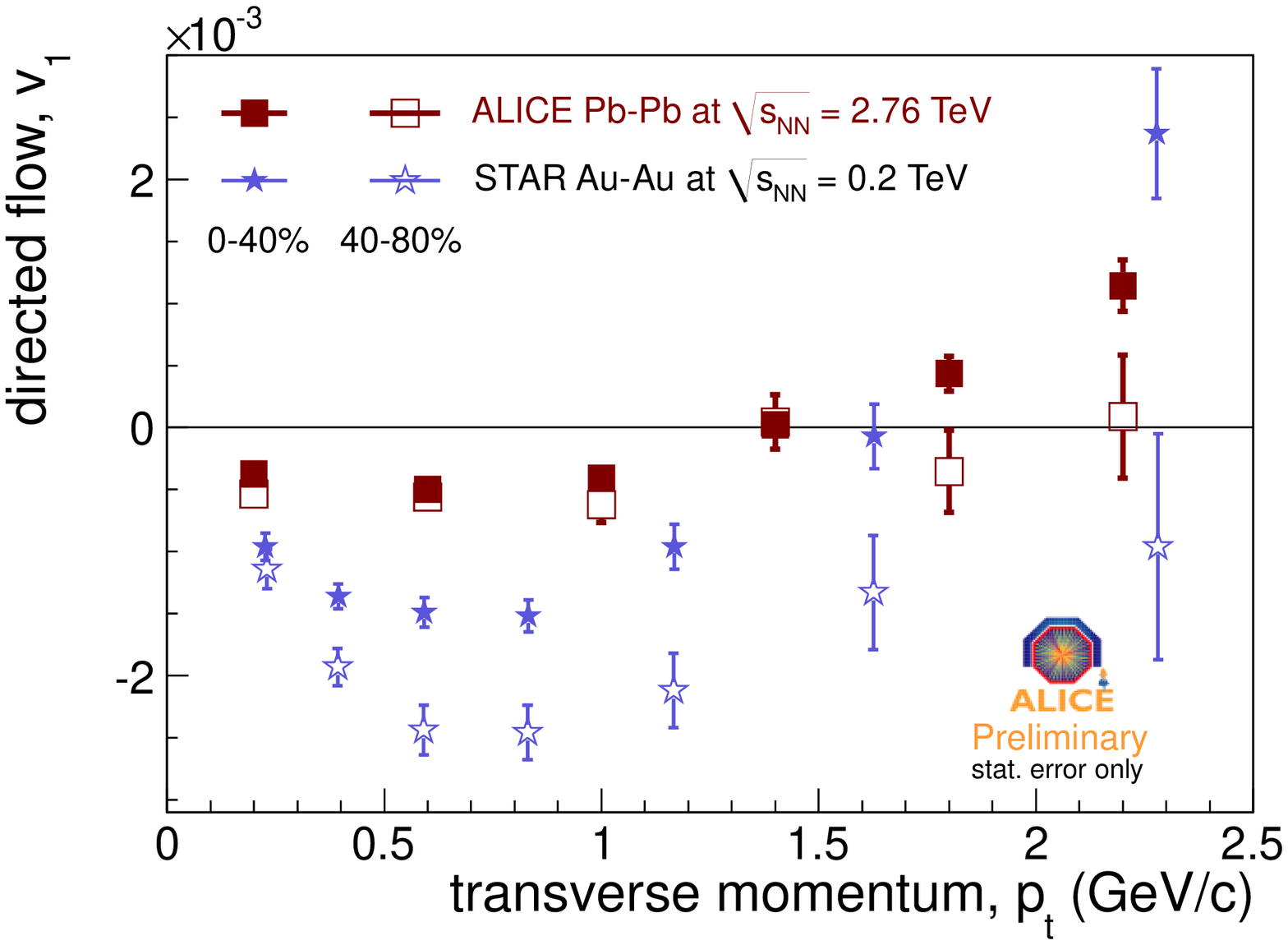}
{\mbox{}\\\vspace{-0.9cm}
\hspace{1cm}\mbox{~} \bf (c)
\hspace{+6.7cm}\mbox{~} \bf (d)}
{\mbox{}\\\vspace{-0.2cm}}
\caption{(color online) (a) $v_1$ over extended rapidity range; (b) longitudinal scaling of $v_1$;
comparison with RHIC: (c) $v_1 (p_{\rm t})$, (d) $v_1$ vs. centrality.}
\label{fig:2}
\end{center}
\end{figure}

\noindent{\bf Results.}~Directed flow over extended rapidity range is shown in Fig.~\ref{fig:2}(a).
We observe that charged particle $v_1(\eta)$ has a negative slope similar to that at RHIC energies. 
$v_1$ shifted to beam rapidity (Fig.~\ref{fig:2}(b)) exhibits
longitudinal scaling previously observed at RHIC \cite{Abelev:2008jga,Back:2005pc}.
Figure~\ref{fig:2}(c) shows that $v_1$ at LHC has a weak centrality dependence and is significantly smaller than at RHIC.
As a function of $p_{\rm t}, $ $v_1$ at RHIC and LHC shows a similar trend (Fig.~\ref{fig:2}(d)),
i.e. sign change around $p_{\rm t} \sim 1.5$ GeV/$c$ in central collisions,
with no zero crossing for peripheral collisions.

\begin{figure}[ht]
\begin{center}
\includegraphics[width=.5\textwidth]{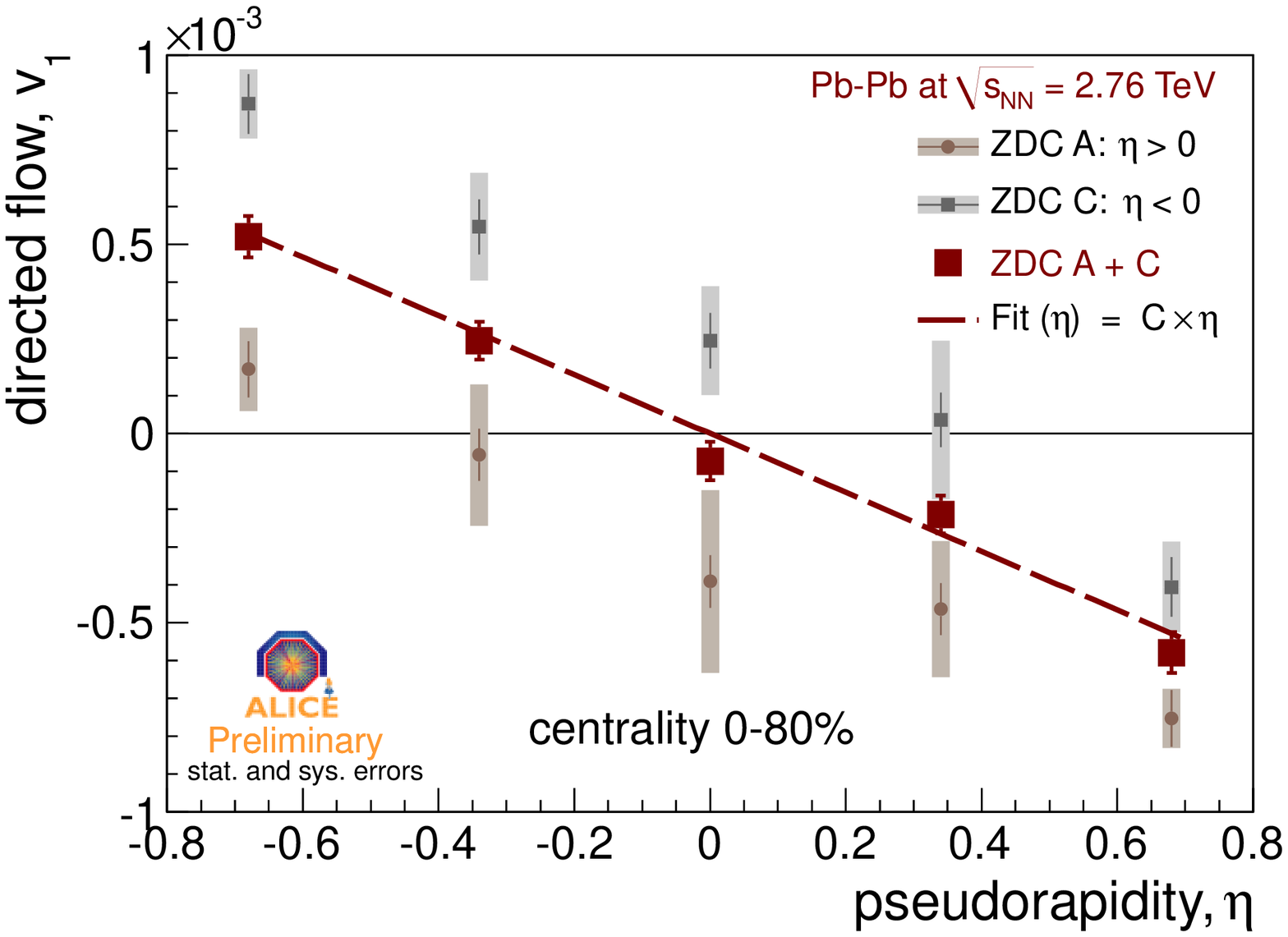}%
\includegraphics[width=.5\textwidth]{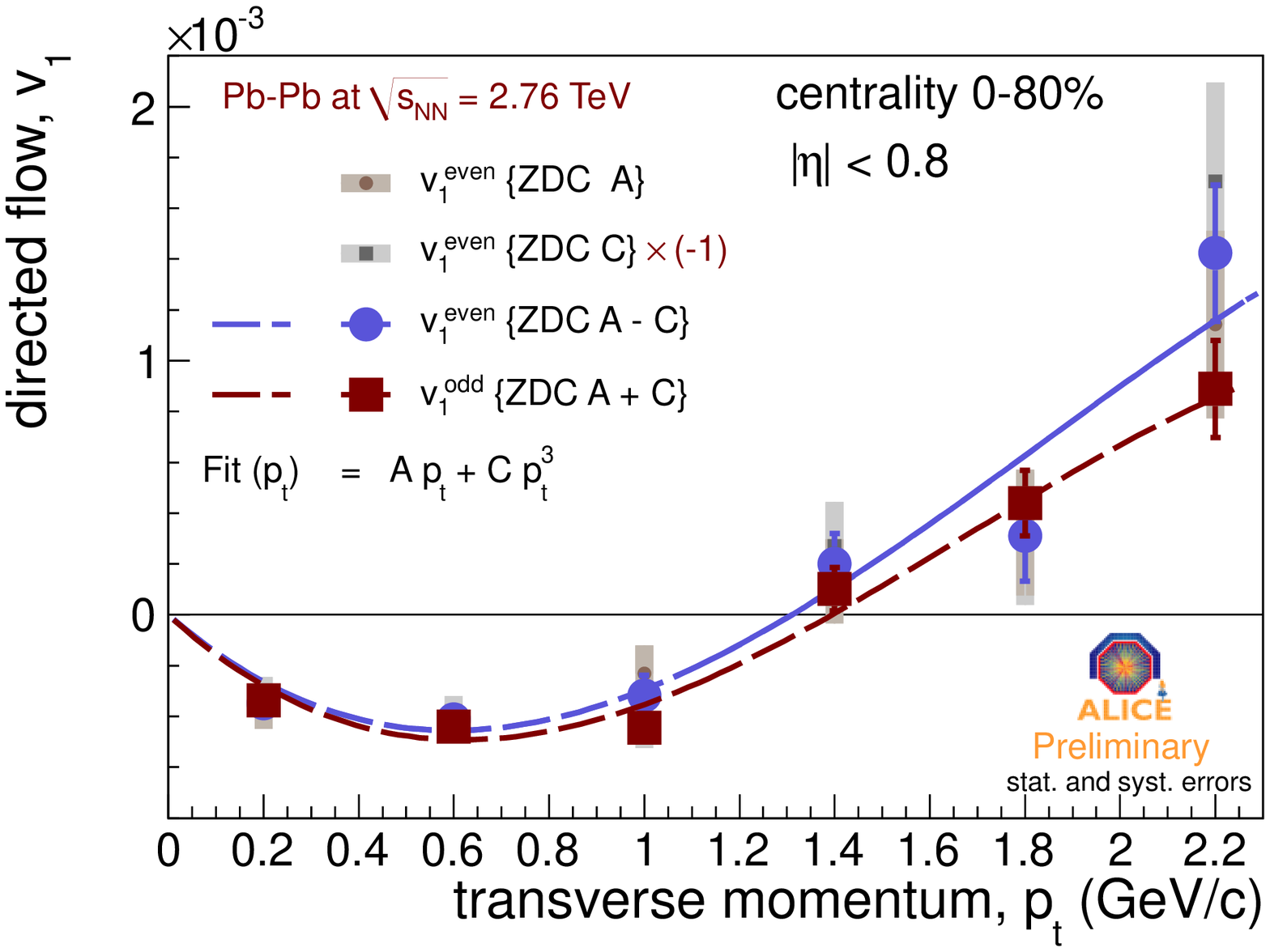}
{\mbox{}\\\vspace{-0.9cm}
\hspace{1cm}\mbox{~} \bf (a)
\hspace{+6.7cm}\mbox{~} \bf (b)}
{\mbox{}\\\vspace{-0.2cm}}
\caption{(color online) Even and odd $v_1$ components: (a) $v_1(\eta)$, (b) $v_1(p_{\rm t})$.}
\label{fig:3}
\end{center}
\end{figure}

\noindent{\bf Flow fluctuations.}~Recent theoretical calculations \cite{Teaney:2010vd,Gardim:2011qn}
predict an $\eta$-even $v_1$ which originates from fluctuations in the initial geometry of the collision.
Experimentally we can separate $\eta$-even and $\eta$-odd $v_1$ by symmetrizing or anti-symmetrizing measured
$v_1^{\rm exp}$ as a function of rapidity, $v_1^{\rm exp}= v_1^{\rm odd}+v_1^{\rm even}$,
\begin{equation}
v_1^{\rm odd}= [v_1^{\rm exp}(\eta)-v_1^{\rm exp}(-\eta)]/2~~,~
v_1^{\rm even}= [v_1^{\rm exp}(\eta)+v_1^{\rm exp}(-\eta)]/2~.
\end{equation}
Asymmetric part, $v_1^{\rm odd}$, reflects correlations with respect to the reaction plane,
while symmetric part, $v_1^{\rm even}$, can be non-zero due to flow fluctuations.
Figure \ref{fig:3}(a) shows
results for $v_1(\eta)$ with A and C side ZDC separately which reveal the symmetric part in $v_1$.
As a function of $p_{\rm t}$ even and odd $v_1$ have a similar shape and magnitude
(Fig.~\ref{fig:3}(b)).
This $p_{\rm t}$ dependence is also very similar
to that of midrapidity $v_1$ extracted from Fourier fits of two particle correlations \cite{Adare:2011qm},
while the magnitude of even and odd $v_1$ measured with spectators is smaller by a factor 40.

\noindent{\bf Summary.}~Directed flow, $v_1$, of charged particles is measured
over a wide range of rapidity, $|\eta| < 5.1$, 
in Pb-Pb collisions at $\sqrt{s_{\rm NN}} = $2.76 TeV.
Magnitude of $v_1$ is about 3-4 times smaller than at top RHIC energy
with a weak centrality dependence.
$v_1(p_{\rm t})$ crosses zero at $p_{\rm t} \sim 1.5$~GeV/$c$.
$v_1(\eta)$ has the same (negative) slope as at RHIC
in contrast to some of the theoretical expectations.
As a function of beam rapidity,
measured $v_1$ is consistent with a picture of
longitudinal scaling observed at RHIC.

Fluctuation of  directed flow of spectators
and particles produced at midrapidity are found to be correlated.
Observed similarity in $p_{\rm t}$ shape of rapidity even and odd
$v_1$ components as well as in $v_1$ extracted from two particle correlations
might indicate a common origin for all three effects.

\end{document}